\documentclass[12pt]{article}
\usepackage[a4paper,top=2.5cm,left=3cm,right=3cm,bottom=2.5cm]{geometry}
\usepackage{url,hyperref}
\usepackage[square,numbers,sort&compress]{natbib}
\usepackage{graphicx,epstopdf,epsfig}
\usepackage[english]{babel}
\usepackage[T1]{fontenc}
\usepackage{verbatim}
\usepackage{listings}
\usepackage{indentfirst}
\usepackage{amssymb,amsmath,amsfonts,mathrsfs,bbm}
\usepackage[bbgreekl]{mathbbol}
\usepackage{color}
\usepackage[font=small,indention=0pt,labelsep=period,margin=1cm]{caption}
\usepackage{times}
\usepackage{soul}
\usepackage{titlesec}
\makeatletter
\def\amsbb{\use@mathgroup \M@U \symAMSb}
\def\@maketitle{\newpage\vspace*{-1cm}{\noindent\bfseries\LARGE{\fontfamily{cmr}\selectfont\@title}\par} \vglue 20pt
  \noindent{\large{\fontfamily{cmr}\selectfont\@author}\par} \vglue 20pt
}
\let\maketitleOLD\maketitle
\renewcommand{\maketitle}{\maketitleOLD\thispagestyle{empty}}
\renewenvironment{abstract}{%
      \list{}{\advance\topsep by6pt\relax
      \leftmargin=1cm \rightmargin=1cm \labelwidth=\z@ \listparindent=\z@ \itemindent\listparindent \rightmargin\leftmargin}
      \item[\hskip\labelsep \bfseries Abstract.]}
      {\endlist}
\makeatother
\titleformat{\section}{\fontfamily{cmr}\selectfont\large\bfseries}{\thesection.}{5pt}{}
\titlespacing{\section}{0pt}{*2}{\wordsep}
\titleformat{\subsection}{\fontfamily{cmr}\selectfont\normalsize\bfseries}{\thesubsection.}{5pt}{}
\titlespacing{\subsection}{0pt}{*1}{\wordsep}

\titlelabel{\thetitle.\hspace{1ex}}
\newcommand*{\doi}[1]{\href{http://dx.doi.org/#1}{doi: #1}}
\newcommand{\bv}{\ensuremath{\boldsymbol{v}}}
\newcommand{\bx}{\ensuremath{\boldsymbol{x}}}
\newcommand{\by}{\ensuremath{\boldsymbol{y}}}
\newcommand{\bk}{\ensuremath{\boldsymbol{k}}}

\newcommand{\bX}{\ensuremath{\boldsymbol{X}}}

\newcommand{\Wt}[1]{\ensuremath{\boldsymbol{W}_{\!\!#1}}}

\newcommand{\dd}{\ensuremath{\mathrm{d}}}
\newcommand{\Rey}{R\hspace{-.5pt}e}
\newcommand{\be}{\begin{equation}}
\newcommand{\ee}{\end{equation}}
\newcommand{\av}[1]{{\left\langle #1 \right\rangle}}
\newcommand{\avnu}[1]{\mathbb{E}^\nu{\left[ #1 \right]}}

\newcommand{\inria}{Universit\'e C\^ote d'Azur, Inria, CNRS\\ Calisto team, Sophia Antipolis, France}
\newcommand{\cemef}{Mines Paris, PSL University, CNRS\\ Cemef, Sophia Antipolis, France}
\newcommand{\inphyni}{Universit\'e C\^ote d'Azur, CNRS\\ Institut de Physique de Nice, France}

\graphicspath{{./figures/}}

\title{Anomalous Dissipation \& Spontaneous Stochasticity in Deterministic Surface Quasi-Geostrophic Flow}
\author{Nicolas Valade, Simon Thalabard, J\'er\'emie Bec}

\begin{document}

\maketitle

\begin{abstract}
Surface quasi geostrophy (SQG) describes the two-dimensional active transport of a temperature field in a strongly stratified and rotating environment. Besides its relevance to geophysics, SQG bears formal resemblance with various flows of interest for turbulence studies, from passive scalar and Burgers to incompressible fluids in two and three dimensions.
This analogy is here substantiated by considering the turbulent SQG regime emerging from deterministic and smooth initial data prescribed by the superposition of a few Fourier modes.  While still unsettled in the inviscid case, the initial value problem is known to be mathematically well-posed when regularised by a small viscosity. In practice, numerics reveal that in the presence of viscosity, a turbulent regime appears in finite time, which features three of the distinctive \emph{anomalies} usually observed in three-dimensional developed turbulence: \textit{(i)} dissipative anomaly, \textit{(ii)} multifractal scaling, and \textit{(iii)} super-diffusive separation of fluid particles, both backward and forward in time. These three anomalies point towards three spontaneously broken symmetries in the vanishing viscosity limit: scale invariance, time reversal and uniqueness of the Lagrangian flow, a fascinating phenomenon that Krzysztof Gaw\k{e}dzki dubbed \emph{spontaneous stochasticity}. In the light of Gaw\k{e}dzki's work on the passive scalar problem, we argue that  spontaneous stochasticity and irreversibility are intertwined in SQG, and provide numerical evidence for this connection. Our numerics, though, reveal that the deterministic SQG setting only features a tempered version of spontaneous stochasticity, characterised in particular by non-universal statistics.
\end{abstract}


\section{Introduction}
\label{sec:introduction}
\noindent Viscous, incompressible fluid velocity fields $\bv(\bx,t)$ solve the Navier--Stokes equations with viscosity $\nu$
\be
	\partial_t \bv + \bv \cdot \nabla \bv  = -\nabla p + \nu \Delta \bv \;, \;\;\;\; \nabla \cdot \bv = 0.
\ee
They evolve into an unsteady, turbulent state when the injection of kinetic energy overwhelms viscous damping. This imbalance is measured by the Reynolds number $\Rey \propto \nu^{-1}$, which in practice can reach very large values, from $10^3$ for the aorta to $10^{12}$ in a cyclone. To efficiently dissipate energy at large Reynolds numbers, the flows develops violent structures at small scales, associated to large fluctuations of velocity gradients. Strikingly, a finite energy dissipation persists when $\Rey\!\to\!\infty$, a phenomenon known as the \emph{dissipative anomaly}. When abruptly setting $\nu=0$, the viscous Navier--Stokes equations turn into the inviscid Euler equations, which are invariant under time reversal $(\bx,t,\bv)\mapsto(\bx,-t,-\bv)$. In physical terms, the persistence of a finite dissipation when $\Rey\!\to\!\infty$ suggests a mechanism of spontaneous symmetry breaking; Such remanent  \emph{time irreversibility}  has long been considered the main source of difficulties in turbulence  modelling.  Mathematically, the dissipative anomaly originates from the intrinsically singular nature of turbulent fields. Kolmogorov's phenomenological arguments~\cite{kolmogorov1941local} suggest that velocity differences over a distance $\ell$ scale as $|\bv(\bx+\boldsymbol{\ell})-\bv(\bx)| \sim |\boldsymbol{\ell}|^h$ with $h\!=\!1/3$, thus implying that relevant fields in turbulence are \emph{rough}, or non-differentiable.  This idea is formalised by the Onsager theorem~\cite{onsager1949statistical,de2017onsager, isett2018proof}, which states that  the value $h\!=\!1/3$ exactly corresponds to the minimal roughness needed to dissipate energy without the help of viscosity. Onsager theorem suggests to define dissipative Euler flows as the natural physically admissible solutions to the Euler equations. Recent work~\cite{delellis2012hprinciple,bardos2014nonuniqueness} shows however that most singular initial data are \emph{wild}, in the sense that they give rise to infinitely many dissipative Euler flows. This suggests that the criterion of energy dissipation alone is not enough  to univocally prescribe turbulent fields from the Euler equations.

A natural set of ideas involves additional anomalies and other broken symmetries of the Euler equations. In particular,  turbulence displays  systematic anomalous deviations to Kolmogorov self-similarity: This phenomenon, dubbed \emph{intermittency}, reflects the breaking of the symmetry of the Euler equation under scale-invariance~\cite{frisch1995turbulence}: $(\bx,t,\bv)\mapsto(\lambda\bx,\lambda^{1-h}t,\lambda^h\bv)$ for $\lambda >0$ and $h \in \mathbb R$. Practical description of intermittency is usually rooted in  Kolmogorov's hypothesis of refined self-similarity (K62)~\cite{kolmogorov1962refinement}, which in its more popular version connects  the scaling of velocity increments to the multi-fractal nature of energy dissipation. The K62 connection between intermittency and time irreversibility is however suggestive only, in the sense that refined self-similarity essentially points towards a mechanism of multiplicative cascades, which in turn can be formulated in an intrinsic fashion, and without relying on the modelling of a dissipation field~\cite{kolmogorov1962refinement,chevillard2019skewed,mailybaev2022hidden}.

\emph{Spontaneous stochasticity} is likely a third constitutive broken symmetry  of Navier--Stokes turbulence, with implications both for the Lagrangian and  Eulerian  flows. Specifically, the Lagrangian flow  $\bX(t\,|\,\bx_0,t_0)$ tracks the positions at time $t>t_0$ of fluid particles located at $\bx_0$ at time $t_0$. One of Krzystof Gaw\k{e}dzki's major contribution to the modelling of turbulent transport is the idea that incompressibility and roughness may conspire together to make this flow non-unique, eventhough the realisation of the velocity field is prescribed~\cite{gawedzki2001turbulent,gawedzki2006simple,gawedzki2008soluble}. Uniqueness of the Lagrangian flow is not, strictly speaking, a symmetry of the Euler equations. Yet, the framework of spontaneous stochasticity gives a probabilistic interpretation strongly reminiscent of the mechanism of spontaneous symmetry breaking. The introduction of a small noise, sent to zero together with a small-scale regularisation, leads to the explosive separation of fluid particles, which   separate in a finite time, no matter how close they initially are. This procedure also allows to build well-defined probability measures over  Lagrangian trajectories. As such,  while it is broken at a deterministic level, the Lagrangian flow is repaired and  well-defined  in  a probabilistic sense. To date, the scenario of  spontaneous stochasticity for Lagrangian trajectories has been well-formalised only for models of advection by random velocities or simplified versions thereof~\cite{bernard1998slow,falkovich2001particles,kupiainen2003nondeterministic,e2000generalized,chaves2003lagrangian,le2002integration,gawedzki2008soluble,drivas2021life}.  For Navier--Stokes turbulence, it is substantiated by consistent numerical and experimental observations of regimes of explosive separations between fluid particles, a phenomenon known as Richardson's super diffusion, and crucial for the  understanding of turbulent mixing~(see~\cite{salazar2009two} for a review).

In principle, the Eulerian flow is a deterministic mapping between initial and final fields over prescribed time interval. Possible breakdowns of Eulerian flows connect to  a conjecture formulated by Lorenz in the late sixties, stating that multiscale fluid flows could have a drastically unpredictable behaviour if their small scales were sufficiently energetic~\cite{lorenz69predictability, palmer2014real}. Spontaneous stochasticity provides a modern perception of this idea: The dynamical evolution of singular velocity fields could be a mathematically ill-posed problem, with a non-continuous dependence on initial conditions, therefore leading to the finite-time separation of initially undistinguishable fields.  This suggests that turbulent velocity fields should perhaps not be treated individually, as is usually the case for partial differential equations in physics, but rather in terms of probability measures.  Recent work shows that the idea of an intrinsic randomness applies both in simplified turbulence models~\cite{mailybaev2016spontaneous, mailybaev2022spontaneous} and in archetypical hydrodynamical instabilities~\cite{biferale2018rayleigh-taylor, thalabard2020butterfly}.  

To date, the complex interplay between time irreversibility, intermittency and spontaneous stochasticity is elucidated ---\,at least partially\,--- in only few turbulent transport problems,  including in particular Burgers or Kraichnan flows.
Burgers dynamics can be interpreted as the pressure-less active transport of a velocity field by itself; The connection between irreversibility, intermittency and spontaneous stochasticity is mediated through the presence of shocks, in which clusters of fluid particles coming from different regions of space and transporting distinct values of the initial velocity field get trapped. In Burgers turbulence, the shocks constitute the only type of singular dissipative structures; they prescribe the intermittency through their statistical distribution, and they provide a breakdown mechanism for the Lagrangian flow~(see, \textit{e.g.},\cite{frisch2001burgulence}).  While the forward flow  $\bX(t\,|\,\bx_0,t_0)$, $t\ge t_0$  is deterministic and univocally prescribes the destinies of fluid particles, the backward flow $\bX(t\,|\,\bx_0,t_0)$, $t\le t_0$ is non-unique, because particles trapped in shocks loose the memory of their past  positions.  Explicit probabilistic constructions of backward Lagrangian flows  can be achieved in terms of backward Markov processes~\cite{eyink2015spontaneous}. As to Kraichnan flow, they describe the transport of a passive temperature field by a prescribed random Gaussian velocity ensemble, with correlations essentially given by 
\be
	\left \langle  | \bv(\bx+\boldsymbol{\ell},t+\tau) - \bv(\bx,t)|^2\right\rangle \propto  |\boldsymbol{\ell}|^{2h}\,\delta (\tau),  \;\;\;\; h\in (0,1),
\ee
namely statistics which are uncorrelated in time and  Kolmogorov-like in space. In Kraichnan flows, Lagrangian particles behave as Markov processes, which map to Brownian trajectories under suitable rescaling.  From the Feynman--Kac representation of partial differential equations, one can explicitly compute their statistics in terms of parabolic boundary-value problems, involving  forward generators $\mathcal{L}_n$ associated to $n$-point motion.  Solving the boundary-value problems requires selecting specific zero-modes of the forward generators. In particular, the zero-modes determine both the anomalous scaling of $n$-point correlation functions, namely the intermittency, and the separation statistics of particles, namely the Lagrangian spontaneous stochasticity~\cite{bernard1998slow,falkovich2001particles,kupiainen2003nondeterministic,chaves2003lagrangian,gawedzki2008soluble}. Strictly speaking, Kraichnan model does not entail any notion of time irreversibility at the level of the velocity field.  Still, fluctuation-dissipation formulas derived from stochastic representation of the temperature field connect Lagrangian spontaneous stochasticity to the dissipation of the scalar energy~\cite{drivas2017lagrangian,gawedzki2008soluble}.

Understanding the complex interplay between the three broken symmetries remains a challenge in the Navier--Stokes case. Unlike Burgers, the topological nature of the dissipative structures is unknown. Besides, it is unclear how to extend the notion of zero mode to non-Markovian and non-linear settings. The purpose of this work is to discuss the case of Surface Quasi-Geostrophic (SQG) flows, which describe the two-dimensional active transport of a temperature scalar field in a strongly stratified and rotating environment~\cite{blumen1978uniform,held1995surface,lapeyre2017surface}. From a fundamental perspective, SQG shares formal analogies with 3D Navier--Stokes, and as such has received much attention both from the mathematical community, on topics related to the development of  singularities, non-uniqueness and the Onsager theorem~\cite{constantin1994formation,constantin1999behavior,buckmaster2019nonuniqueness}, and from the physical community on topics related to cascades, intermittency,  transport and predictability~\cite{held1995surface,smith2002turbulent,celani2004active,rotunno2008generalization,foussard2017relative,lapeyre2017surface}. We here specifically discuss the turbulent state emerging from a deterministic initial condition, prescribed by a few large-scale Fourier modes, and for which viscous regularisation ensures global existence of the solution. In practice, numerics reveal the emergence of a turbulent state in finite-time, which shares the three broken symmetries of Navier--Stokes turbulence. Due to the specific features of SQG transport, the bridging mechanisms also share features with both Burgers and Kraichnan flows.  Similar to Kraichnan flows, the spontaneous stochasticity is evidenced both for forward and backward trajectories. Similar to Burgers flows, breakdown of the Lagrangian flow connects to anomalous dissipation of the advecting field. Our numerics however point towards the fact that SQG stochasticity might be present in a \emph{tempered} version only, characterised in particular  by non-universal statistics: This  means that the statistics obtained in the double limit of vanishing viscosity and perturbation might be highly sensitive to the way the latter is taken.

The paper is organised as follows. Section~\ref{sec:active} introduces SQG flows,  focusing on  formal analogies and differences  with various flows of fundamental interest, from Burgers and passive scalar to 2D and 3D Navier--Stokes flows.  We discuss freely evolving SQG from analytical initial conditions, and argue from numerics that a turbulent regime emerges, which realises the scenario of an active scalar \textit{\`a la} Kolmogorov. Section~\ref{sec:anomalous} focuses on spontaneous stochasticity. We discuss the connections between Lagrangian backward stochasticity and anomalous dissipation that come from incompressibility and the stochastic representation of advection. We then provide  numerical evidence for Lagrangian stochasticity, both forward and backward in time. In particular, varying the Reynolds number, we evidence a persistent Richardson superdiffusion, with strong correlations to the statistics of the dissipation. We highlight the fact that, unlike in 3D Navier--Stokes, the observed SQG superdiffusion retains dependence upon the initial separation, hereby signalling  tempered  stochasticity. Section~\ref{sec:conclusions} draws concluding remarks, and suggests possible scenarios for the tempered nature of SQG stochasticity.


\section{An active scalar \`a la Kolmogorov}
\label{sec:active}

\subsection{A two-dimensional analog of three-dimensional Navier--Stokes turbulence}
\label{ssec:analog}
\noindent SQG dynamics is described in terms of a surface temperature field $\theta(\bx,t)$ that solves
\begin{equation}
\partial_t \theta + \bv\cdot\boldsymbol{\nabla} \theta = \nu\Delta\theta,\quad\mbox{with } \bv = (-\partial_2\Psi,\partial_1\Psi)^{\!\top} \mbox{ and } |\Delta|^{1/2} \Psi = \theta,
\label{eq:sqg}
\end{equation}
which is nothing but the transport and diffusion of an active scalar. The advecting incompressible velocity $\bv$ is itself controlled by the temperature through a functional relation $\bv = \mathcal{R}^\perp\theta$. This non-local relation simply corresponds to the Fourier multiplier  $\hat{\bv}_{\bk} = i (\bk/|\bk|)^\perp\,\hat{\theta}_{\bk}$, where $\bk^\perp = (-k_2,k_1)^{\!\top}$ and $\hat{\bv}_{\bk}$ and $\hat{\theta}_{\bk}$ are the Fourier transforms of velocity and temperature associated to wavenumber $\bk$, leading to straightforward numerical simulations by pseudo-spectral methods.  Because the $\mathcal{R}^\perp$ operator is homogenous of degree $0$ in $\boldsymbol{k}$-space, the active surface temperature $\theta$ and the advecting velocity $\bv$ have the same dimension and share the same power spectrum. This leads to draw a formal analogy between SQG and Burgers flow, which can be seen as a 1D version of Equation~(\ref{eq:sqg}).

SQG dynamics has actually much in common with the transport of vorticity in 2D Navier--Stokes flows. In particular, SQG flows also have two inviscid quadratic invariants, which we denote as 
\begin{align}
\tag{Hamiltonian}
&\mathcal{H}(t) := \frac{1}{2} \left\langle \theta\,\Psi\right\rangle = \frac{1}{2} \sum_{\bk}  |\bk|^{-1} |\hat{\theta}_{\bk}(t)|^2\\
\tag{Surface kinetic energy}
\text{and}\quad & \mathcal{E}(t) := \frac{1}{2} \left\langle \theta^2\right\rangle  = \frac{1}{2} \left\langle |\bv|^2 \right\rangle = 
\frac{1}{2} \sum_{\bk} |\hat{\theta}_{\bk}(t)|^2,
\end{align}
where the angular brackets $\langle\cdot\rangle$ stand for spatial averages over $[0,2\pi]^2$. They play roles similar to energy and enstrophy in 2D Navier--Stokes turbulence~\cite{held1995surface}: In a developed turbulent regime, the Hamiltonian carries out an inverse cascade towards large scales, while the surface kinetic energy flows down to small scales where it is dissipated by large gradients of $\theta$. Analogies with 2D hydrodynamical turbulence are however at odds on one crucial aspect. While the 2D direct cascade of enstrophy involves a smooth, differentiable flow, the cascade of surface kinetic energy implies rough, non-differentiable velocity fields. Kolmogorov's phenomenology predicts for instance that in SQG turbulence, the temperature field, as well as the velocity, have spatial increments that scale as $\delta_\ell\theta = |\theta(x+\ell)-\theta(x)|\sim\ell^{1/3}$, similarly to 3D Navier--Stokes.  However, direct numerical simulations of SQG in the presence of forcing find that the power spectrum $E(k)$ of the surface temperature is steeper than the Kolmogorov prediction $k^{-5/3}$(see~\cite{lapeyre2017surface} for a review): This is likely a signature that, as in 3D Navier--Stokes, SQG turbulence features intermittency~\cite{sukhatme2002surface,celani2004active}.

\medskip
The analogies between SQG and 3D Navier--Stokes have motivated over the last decades a considerable interest among mathematicians. Following the seminal work of \citet{constantin1994formation}, attention has predominantly focused on the well-posedness of the initial-value problem in the inviscid case $\nu=0$.  Although local existence and uniqueness of regular solutions are granted, the possibility of finite-time singularity from smooth initial data remains open ---\,see~\cite{kiselev2020small} for a recent review.  In the scenario originally proposed in~\cite{constantin1994formation}, the development of a singularity would require the presence of hyperbolic point in the initial level sets of the scalar.  This scenario has been ruled out~\cite{ohkitani1997inviscid,constantin1998nonsingular,cordoba1998nonexistence}, but an alternative framework where patches of surface temperature undergo a self-similar cascade of shear instabilities, gives strong evidence in favour of finite-time blowup~\cite{scott2019scale}.

Besides quests for singularities, results have been obtained on the weak formulation of SQG dynamics~(\ref{eq:sqg}). Global existence of weak solutions with bounded surface kinetic energy is proven in~\cite{resnick1995dynamical}. However, convex integration techniques, which are proved successful for the Euler equation, face here difficulties because the Fourier multiplier relating $\bv$ and $\theta$ is an odd function of wavenumbers~\cite{isett2015holder}. Attention has mostly focused on solutions that violate the conservation of the Hamiltonian $\mathcal{H}(t)$. The SQG version of Onsager's theorem is the conjecture that $\mathcal{H}$ is dissipated if and only if $\delta_\ell\Psi\sim\ell^{1+h}$ with $h\le 0$. While, it was shown in~\cite{isett2015holder} that solutions with $h>0$ conserve the Hamiltonian,  singular flows that dissipate $\mathcal{H}$ can only be explicitly constructed  when $h<-1/5$, at least to this day ~\cite{buckmaster2019nonuniqueness}. If one has in mind, though, a cascade-like mechanism towards the small scales, we expect the surface kinetic energy $\mathcal{E}$ rather than the Hamiltonian $\mathcal{H}$ to be anomalous, so that the dissipation rate $\varepsilon_\theta = \nu \left\langle|\boldsymbol{\nabla}\theta|^2\right\rangle$ has a finite limit when $\nu\to 0$. As for 3D Euler, this should require $\delta_\ell\theta\sim\ell^h$ with $h\le 1/3$ but, to our knowledge, only the necessary condition has be proven for this Onsager-like criterion~\cite{akramov2019renormalization}. Furthermore, the question of uniqueness or non-uniqueness for $h\le 1/3$ remains unsettled, even if the rigidity results of~\cite{isett2015holder} indicate that the vanishing viscosity limit could be well posed. From a mathematical viewpoint, it is thus today unclear whether SQG dynamics is expected or not to display Eulerian spontaneous stochasticity. Still, numerical simulations of SQG dynamics in the presence of forcing suggest the presence of an inverse cascade of errors \cite{rotunno2008generalization,palmer2014real}.

Lagrangian spontaneous stochasticity is nevertheless likely to underpin the separation of tracers in SQG flows.  The numerical simulations of~\cite{smith2002turbulent,foussard2017relative} demonstrate that pairs separate in a superdiffusive manner. The average squared distance between tracers follows approximately Richardson's law $R^2 \propto t^3$ at large-enough times, in agreement with dimensional expectations. Nevertheless, dependence upon either the initial separation $R(0)$ or viscosity has not been fully sorted out. The explosive nature of Lagrangian dispersion thus remains largely unsettled.

The considerations above and related open questions provide motivation for revisiting SQG flows in the light of its broken symmetries, in particular time irreversibility and spontaneous stochasticity. This work aims to be a first step in this direction. To stay as close as possible to the mathematical problems, we restrict our study to SQG flows freely evolving from deterministic smooth initial conditions.

\subsection{Turbulence from analytical initial conditions}
\label{ssec:IC}
\begin{figure}[h]
\centering
\includegraphics[width=\linewidth]{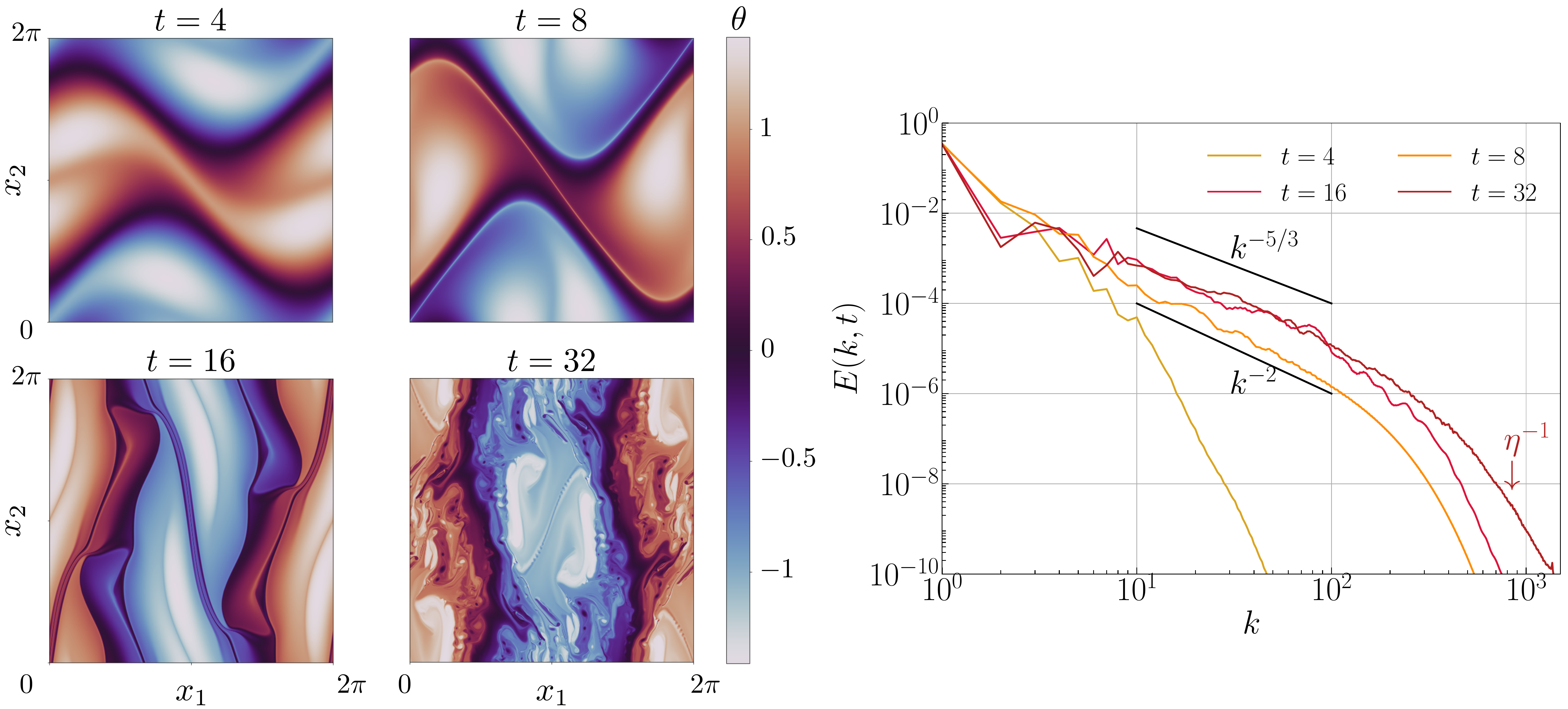}
\caption{\label{fig:snapshots_theta_times} {\it Left:} Snapshots of the surface temperature $\theta$ for $\nu=10^{-5}$ (Run IV), at times before ($t=4,\,8,\,16$) and after ($t=32$) the turbulent regime has settled. {\it Right:} Shell-averaged power spectrum of the surface temperature field $E(k,t)=\frac{1}{2} \sum_{k-1<|\bk|\le k} |\hat\theta_{\bk}(t)|^2$ computed at the same instants of time. The solid lines correspond to behaviours $\propto k^{-2}$ and $\propto k^{-5/3}$.}
\end{figure}

\noindent We consider decaying solutions to the SQG equation~(\ref{eq:sqg}), whose initial condition at time $t=0$ is smooth, analytic and consists of the superposition of a few Fourier modes. The idea is to highlight that, even if the question of a finite-time blowup remains open for $\nu=0$, such solutions display turbulent features in the asymptotics of small viscosities. More specifically, we consider the initial condition introduced by \citet{constantin1994formation}
\begin{equation}
	\theta(\bx,0) = \cos x_2 - \sin x_1\,\sin x_2. 
	\label{eq:int_cond_CMT}
\end{equation}
The surface kinetic energy at time $t=0$ is $\mathcal{E} = 3/8$, so that the typical time and velocity scales are of the order of unity. Numerical simulations are performed using a fully-dealiased pseudo-spectral method at various resolutions and values of the viscosity (see Tab.~\ref{tab:num}). As seen in the left panels of Fig.~\ref{fig:snapshots_theta_times}, the solution here shown for Run IV, starts smooth ($t=4$), develops a quasi-singular filament ($t=8$) that continues to get stretched and folded by the flow ($t=16$). At a large-enough time ($t=32$), these filaments have destabilised, activating a wide range of scales in the active scalar field.

The corresponding power spectra are shown in the right panel of Fig.~\ref{fig:snapshots_theta_times}. The development of a quasi-singular behaviour at time $t\approx 8$ goes with the development of an intermediate range where $E(k)$ approximately scales as $k^{-2}$, which is consistent with the measurements of \citet{ohkitani1997inviscid}. This scaling can be explained as the signature of discontinuities of the temperature field $\theta$ occurring on lines, as visible from Fig.~\ref{fig:snapshots_theta_times}b. At larger times, the solution experiences a series of quasi-singular episodes. As visible in the spectrum at time $t=16$, they materialise as waves that propagate towards large $k$'s.  The solution's power spectrum gets gradually less steep and approach the behaviour closer to $k^{-5/3}$ expected in the developed turbulent regime.

\begin{table}[t]
	\begin{center}
	\begin{tabular}{ ||c||c|c|c|c|c|| } 
 		\hline
	 	Run & I & II & III& IV & V \\ \hline
 		$\nu$ & $3\times10^{-4}$ & $10^{-4}$ & $3\times10^{-5}$ & $10^{-5}$ & $3\times10^{-6}$\\ 
 		$N^2$ & $1024^2$ & $1024^2$ & $2048^2$ & $4096^2$ & $8192^2$\\ 
		$\Delta t$ & $4\times10^{-4}$ & $4\times10^{-4}$ & $2\times10^{-4}$ & $10^{-4}$ & $5\times10^{-5}$ \\
		$\eta(t=32)$  & $1.7\times10^{-2}$& $7.7\times10^{-3}$& $3\times10^{-3}$& $1.3\times10^{-3}$& $5.3\times10^{-4}$\\ \hline
	\end{tabular}
	\end{center}
\caption{\label{tab:num} Parameters of the different numerical simulations: viscosity $\nu$, number of collocation points $N^2$, time step $\Delta t$, dissipative scale $\eta = (\nu^{3}/\langle\varepsilon_\theta\rangle)^{1/4}$ at time $t=32$, where $\langle\varepsilon_\theta\rangle$ is the spatially averaged dissipation rate of surface kinetic energy, as defined in~(\ref{eq:dissip_rates}). Each run additionally contains $N_{\rm tr} = N^2$ Lagrangian particles. }
\end{table}

\begin{figure}[h]
\centering\includegraphics[width=.75\linewidth]{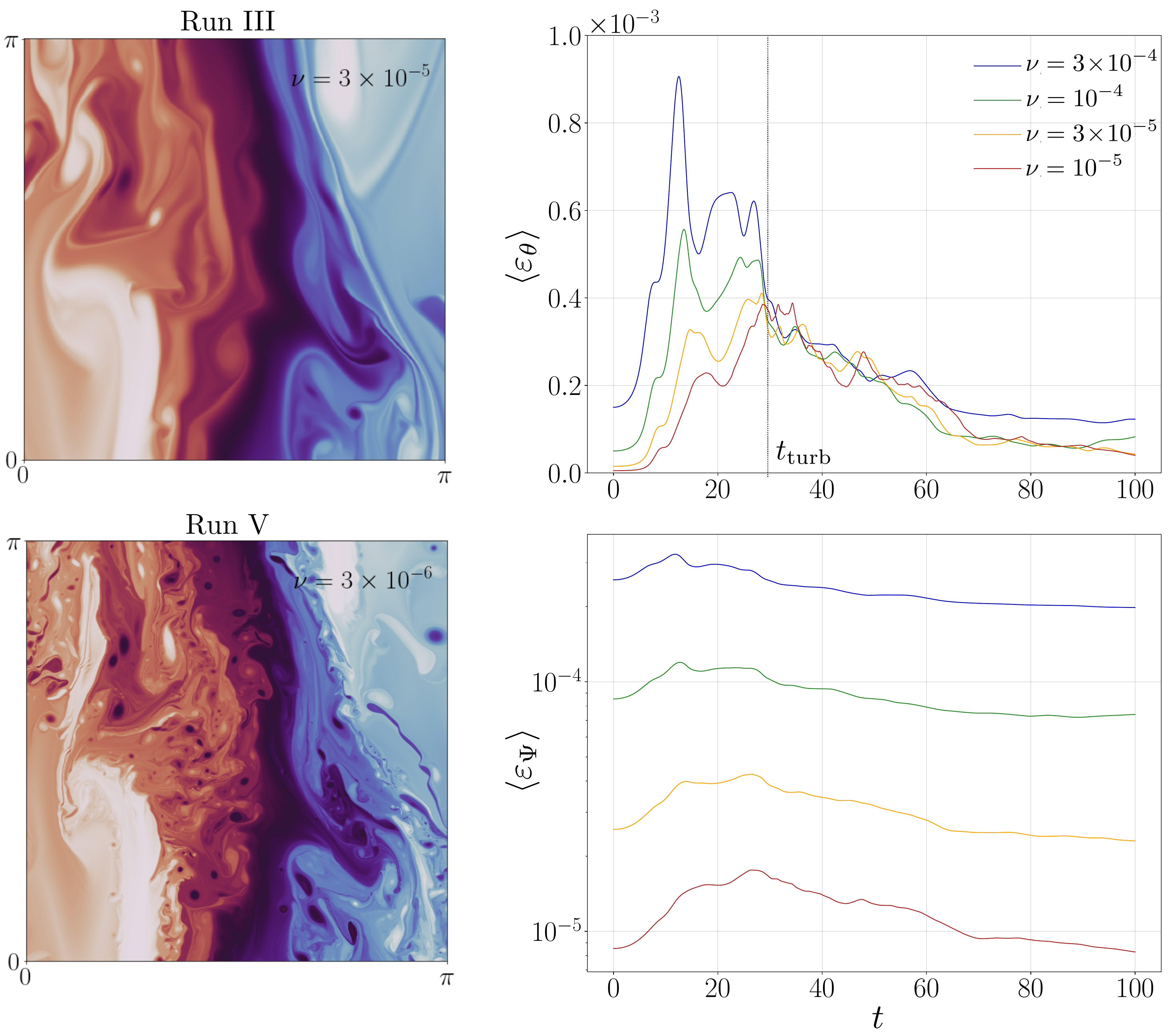}
\caption{\label{fig:dissip_anomal} {\it Left:} Snapshots at time $t=32$ of $\theta$ in the quadrant $[0,\pi]^2$ for $\nu=3\times10^{-5}$ (top) and $\nu=3\times10^{-6}$ (bottom); The colour scale is the same as in Fig.~\ref{fig:snapshots_theta_times}. {\it Right:} Spatially averaged dissipation rates $\langle\varepsilon_{\Psi}\rangle$ of the Hamiltonian $\mathcal{H}$ (up) and $\langle\varepsilon_{\theta}\rangle$ of the surface energy $\mathcal{E}$ (bottom), as a function of time for the different values of the viscosity $\nu$.}
\end{figure}
We now turn to dependence of the turbulent regime upon the viscosity $\nu$.  The right panels of Fig.~\ref{fig:dissip_anomal} display snapshots of the solution at time $t=32$ for two different viscosities. One clearly observes that decreasing viscosity leads to the formation of thiner filaments that are more likely to destabilise and create structures at even smaller scales. This picture agrees with the scenario of cascading shear instabilities studied in~\cite{scott2019scale}. The left panels of Fig.~\ref{fig:dissip_anomal} show for the various $\nu$, the time evolutions of the dissipation rates
\be
	\langle\varepsilon_{\Psi}\rangle  = -\frac{{\rm d}\mathcal{H}}{{\rm d} t} = \nu \left\langle \nabla\Psi\cdot\nabla\theta \right\rangle \quad\mbox{and}\quad \langle\varepsilon_{\theta}\rangle = -\frac{{\rm d}\mathcal{E}}{{\rm d} t} = \nu \left\langle |\nabla\theta|^2\right\rangle,
	\label{eq:dissip_rates}
\ee
where, as before, $\langle\cdot\rangle$ denotes spatial average. Measurements indicate that the dissipation of $\mathcal{H}$ seems to decrease proportionally to $\nu$ when we approach the inviscid limit, indicating the absence of dissipative anomaly for the Hamiltonian.  Conversely, the dissipations of $\mathcal{E}$ associated to different viscosities collapse to a behaviour independent of $\nu$ at large-enough times. This gives evidence that an anomalous dissipation of the surface kinetic energy could persist in the limit $\nu\to0$.  Note that the time needed for such a regime to establish seems independent of the viscosity, at least in the range of values that is considered here. Indeed, all curves on the bottom-right panel of Fig.~\ref{fig:dissip_anomal} are pinched at a time $t_{\rm turb}\approx 27$ --- $30$. This means that either \textit{(i)}~dependence of this timescale on viscosity is very weak (almost invisible over two decades), or \textit{(ii)}~a specific singular event occurs at this time in the inviscid solution. The first option could for instance result from a double exponential growth of temperature gradients~\cite{ohkitani1997inviscid,constantin1998nonsingular}, which would lead to the double-logarithmic behaviour $t_{\rm turb} \sim \log( |\log \nu|)$, impossible to detect from our simulations. The second option might involve the hurling cascade of shear instabilities of~\cite{scott2019scale}, but before the solution has to cross several almost-singular events (illustrated for instance in Fig~\ref{fig:snapshots_theta_times}b and c), where temperature gradients become so large that even very small viscosities are brought into play. Such an eventuality could be a hindrance to detect any blowup with current computational means.

\begin{figure}[th]
\centering\includegraphics[width=\linewidth]{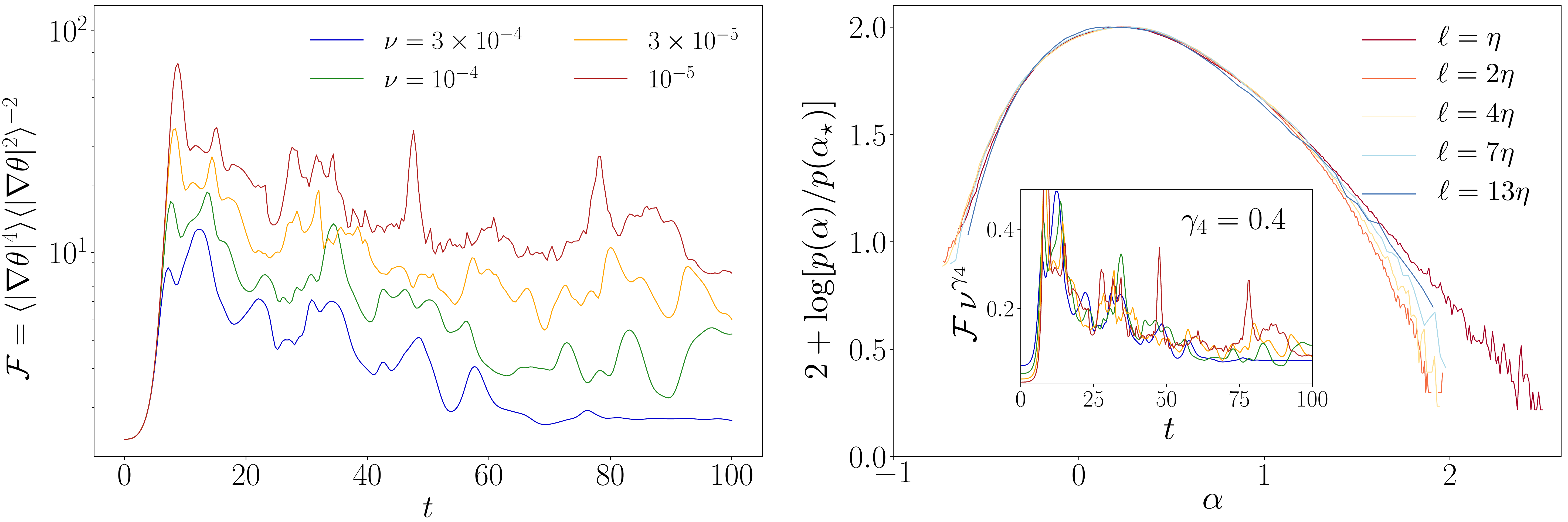}
\caption{\label{fig:intermittency} {\it Left:} Time evolution of the flatness $\mathcal{F}$ of temperature gradients $\nabla\theta$ (or equivalently of the normalised variance of $\varepsilon_\theta$) for different viscosities.  {\it Right:} Probability density functions $p(\alpha)$ of the local scaling exponent $\alpha :=  \log \left(\langle \varepsilon_\theta\rangle_\ell/\langle \varepsilon_\theta\rangle\right) / \log(\ell/L)$ of the coarse-grained dissipation $\langle\varepsilon_\theta\rangle_\ell$ for $\nu=10^{-5}$ (Run IV) and over various scales $\ell$ as labelled. Here $p(\alpha_\star) = \sup_\alpha p(\alpha)$ and we have chosen $L = 2\pi$. {\it Inset:} same as the left-hand panel, but rescaling this time $\mathcal{F}$ with $\nu^{-\gamma_4}$ with here $\gamma_4 = 0.4$.}
\end{figure}

Regardless of the presence or not of finite-time singularities, a turbulent regime settles at large-enough times $t>t_{\rm turb}$. Remarkably, the decaying solution develops features that are very similar to those observed for statistically stationary solutions sustained by an external forcing. Besides power-law behaviours of energy spectra, the solution displays features expected for an intermittent turbulent field. The left-hand panel of Fig.~\ref{fig:intermittency} shows the time evolutions of the normalised fourth-order moment $\mathcal{F} := \left\langle|\nabla\theta|^4\right\rangle/\left\langle|\nabla\theta|^2\right\rangle^2$ of the temperature. The curves associated to different values of the viscosity remain approximately parallel to each other, indicating that gradients feature an anomalous scaling as a function of $\nu$, as known to occur in 3D Navier--Stokes turbulence~\cite{nelkin1990multifractal}.

To characterise further $\varepsilon_\theta$,  we have also studied the distribution of the coarse-grained dissipation $\left\langle \varepsilon_\theta\right\rangle_\ell (\bx,t) := (1/\pi\ell^2)\int_{|\boldsymbol{r}|<\ell/2} \varepsilon_\theta(\bx+\boldsymbol{r},t){\rm d}^2r$.   Turbulent dissipation fields are expected to strongly deviate from self-similarity and to rather display multifractal statistics (see, \textit{e.g.}, \cite{frisch1995turbulence}). In particular this means that the coarse-grained dissipation scales as $\langle \varepsilon_\theta \rangle_\ell/ \langle \varepsilon_\theta \rangle  \sim (\ell/L)^\alpha$ on a fractal set of dimension $\mathcal{D}(\alpha)$, \textit{i.e.}\/ with a probability $\propto (\ell/L)^{2-\mathcal{D}(\alpha)}$ . As seen on the right-hand side of Fig.~\ref{fig:intermittency}, the distributions of the singularity exponent $\alpha$ obtained for various scales $\ell$ collapse on the top of each other. This is in agreement with multifractal statistics and the resulting master curve provides an approximation of $\mathcal{D}(\alpha)$. It is worth mentioning that estimating these distributions required, in addition to spatial averages, a time average over $t\in[40,50]$.  Similarly to Navier--Stokes turbulence~\cite{nelkin1990multifractal,frisch1995turbulence}, multifractal distributions of the coarse-grained dissipation can be used to predict through saddle-point arguments how the moments of the temperature gradients scale as a function of viscosity. Their flatness is for instance expected to behave for $\nu\to0$ as $\mathcal{F} \sim \nu^{-\gamma_4}$ where $\gamma_4 = \sup_\alpha [3(\mathcal{D}(\alpha)-2-2\alpha)/(4+\alpha)]$. The approximation of $\mathcal{D}(\alpha)$ obtained numerically leads to $\gamma_4 = 0.4 \pm 0.05$, which is consistent with our data (see the inset of the right panel of Fig.~\ref{fig:intermittency}). These results on the intermittency of the active scalar $\theta$ further support analogies between SQG and 3D Navier--Stokes, which share a clear break-up of scale-invariance symmetries.


\section{Anomalous dissipation and spontaneous stochasticity}
\label{sec:anomalous}
\subsection{From fields to trajectories}
\label{ssec:fieldstotraj}
\noindent We now turn specifically to the issue of spontaneous stochasticity. For SQG, the presence of anomalous dissipation implies the breakdown of Lagrangian flows, both backward and forward in time.  This connection essentially stems from the stochastic representation of  advection,  involving backward-in-time statistical averages over tracers~\cite{gawedzki2008soluble,constantin2008stochastic}, and which applies to both passive and active transport~\cite{drivas2017lagrangian}. The feature specific to SQG is the fact that the scalar dissipation identifies to the energy dissipation.

Following the terminology of \cite{gawedzki2008soluble}, tracers designate noisy Lagrangian fluid particles. Specifically they solve for $s\!\ge\!0$ the stochastic differential equations
\begin{equation}
	\label{eq:tracers}
	\dd\bX(s\,|\,\bx,t) = \bv\!\left(\bX(s\,|\,\bx,t),s\right)\mathrm{d}s + \sqrt{2\nu}\,\dd \Wt{s},\;\;\;\;\; \bX(t\,|\,\bx,t)=\bx,
\end{equation}
where $\Wt{s}$ denotes the 2D Brownian motion. This evolution defines for $s\ge t$, \textit{forward} trajectories emanating from  $\bx$ at time $t$, and for $s \le t$, \textit{backward} trajectories winding to $\bx$ at time $t$. In lieu of the Lagrangian flow,  tracers determine the transition probabilities
\be
	\label{eq:tran}
	p^\nu\!(\by,s\,|\,\bx,t;\theta_0)\,\dd y_1\dd y_2 :=  \mathbb{P}^\nu\!\left[ \bX(s\,|\,\bx,t)\in [y_1,y_1\!+\!\dd y_1]\!\times\![y_2, y_2\!+\!\dd y_2] \right].
\ee
Here and in the sequel, $\mathbb{P}^\nu$ and $\mathbb{E}^\nu$ denote probability and expectation with respect to the Brownian motion $\Wt{s}$.  These probability densities are defined both forward (for $s\ge t$) and backward (for $s\le t$). Owing to incompressibility, forward and the backward densities relate to each to other through
\be
	\label{eq:BF}
	p^\nu\!(\by,s\,|\,\bx,t;\theta_0) = p^\nu\!(\bx,t\,|\,\by, s;\theta_0).
\ee
Itô calculus \cite{evans2012introduction} provides the stochastic representation for the temperature field in terms of backward averaging 
\be
	\label{eq:av}
	\theta(\bx,t) = \avnu{\theta_0(\bX(0\,|\,\bx,t))} = \int\theta_0(\by)\,p^\nu\!(\by,0\,|\,\bx,t;\theta_0)\,\dd^2 y,
\ee
 together with the fluctuation estimate
 \be
	\label{eq:fluct_int}
	 \avnu{\left[\theta(\bx,t) - \theta_0(\bX(0\,|\,\bx,t))\right]^2} =  
		  2 \nu \int_0^t \avnu{|\nabla \theta(\bX(s\,|\,\bx,t),s)|^2} \dd s
\ee
Expanding the left-hand side of (\ref{eq:fluct_int}) using  both the representation (\ref{eq:av}) and the incompressibility condition (\ref{eq:BF}), one obtains the following stochastic representation for the dissipation of scalar variance, or equivalently of the surface kinetic energy 
\be
	\label{eq:fluct}
	\mathcal E(0) -\mathcal E(t) = \int_0^t\!\! \varepsilon_\theta(s)\,\dd s = \dfrac{1}{8\pi^2}\!\iint\!p^\nu\!(\by,0\,|\,\bx,t;\theta_0) \;{\left(\theta(\bx,t)-\theta_0(\by) \right)^2} \dd^2 x\,\dd^2 y.
\ee
This relation is the SQG-version of a kinematic criterion highlighted by Gaw\c{e}dzki in his lectures notes on the passive scalar problem~\cite{gawedzki2008soluble}. As pointed out in~\cite{drivas2017lagrangian} for more general settings, Equation~(\ref{eq:fluct}) is a fluctuation-dissipation relation, which ties the dissipation of an Eulerian inviscid invariant to the fluctuations of the temperature field along the Lagrangian paths. The presence of a dissipative anomaly implies that the transition probability is not concentrated on deterministic paths. In the vanishing viscosity limit, where the tracers (\ref{eq:tracers}) formally become deterministic Lagrangian particles, the  persistence of a kinematic dissipation on a finite time interval therefore implies the breakdown of the backward Lagrangian flow. In that sense, this mechanism ressembles the Burgers phenomenology recalled in the introduction. Yet, there is a fundamental difference: From incompressibility and the relation (\ref{eq:BF}), the breakdown of the backward Lagrangian flow implies that the forward Lagrangian flow also breaks down.

\subsection{Numerical evidence for Lagrangian spontaneous stochasticity}
\label{ssec:spontaneous}
\noindent From Eq.~(\ref{eq:fluct}), we inferred that the time irreversibility of SQG, measured at the level of the surface kinetic energy, implies the presence of Lagrangian stochasticity, both forward and backward in time.  For the turbulent regime discussed  in\S\ref{ssec:IC}, the apparent collapse from $t_{\rm turb} \approx 30$ for the evolution of the  dissipation of Fig.~\ref{fig:dissip_anomal} suggests finite-time emergence of anomalous dissipation in the limit $\nu\to 0$.  Although in an indirect manner, this points towards the presence of non-deterministic trajectories in the inviscid limit. To substantiate this possibility from direct numerical observations, we estimate the transition probabilities $p^\nu$  given by Eq.~(\ref{eq:tran}) using Monte-Carlo sampling of Lagrangian particles seeded in the flow. We use statistical averages over puffs of particles extended over a few $\eta$  to estimate averages over the Brownian noise. In other words, we use the identification
\be  
	\label{eq:mc}
	\avnu{\cdot} \simeq \av{\cdot}_{\ell_\nu},
\ee
with $\ell_\nu$ prescribed of the order of $\eta\!=\!(\nu^3/\varepsilon_\theta)^{1/4}$. 
We evaluate the relative separation of pairs of tracers that coincide in $\bx$ at time $t$, both forward and backward in time, as
\be
	\av{R^2(\pm \tau\,|\,\bx,t)}_{\ell_\nu} := \av{\left|\bX(t\pm\tau\,|\,\bx,t) - \bX'(t\pm\tau\,|\,\bx,t)\right|^2}_{\ell_\nu}
\ee
with $\bX$ and $\bX'$ solutions associated to independent realisations of the Brownian motion in Eq.~(\ref{eq:tracers}).
In principle, this quantity relates to the transition probability $p_\nu$ through
\begin{eqnarray}
	 \av{R^2(\pm \tau\,|\,\bx,t)}_{\ell_\nu} &\simeq& \iint |\by-\by'|^2\,p^\nu\!(\by,t\pm\tau\,|\,\bx,t)\,p^\nu\!(\by',t\pm\tau\,|\,\bx,t) \,\dd^2y\,\dd^2y' \nonumber\\
	 &\simeq& 2 \left(\avnu{\left|\bX(t\pm\tau\,|\,\bx,t)\right|^2}- \left|\avnu{\bX(t\pm\tau\,|\,\bx,t)}\right|^2\right)\!.
\end{eqnarray}
This shows that the relative separation simply prescribes the variance of the tracers starting or ending in $\bx$ at time $t$. Hence, the persistence of non-zero values of $\av{R^2(\pm \tau\,|\,\bx,t)}_{\ell_\nu}$ when $\nu\to0$ signals non-deter\-mi\-nistic transition probabilities in the inviscid limit.

\begin{figure}[h]
\includegraphics[width=\linewidth]{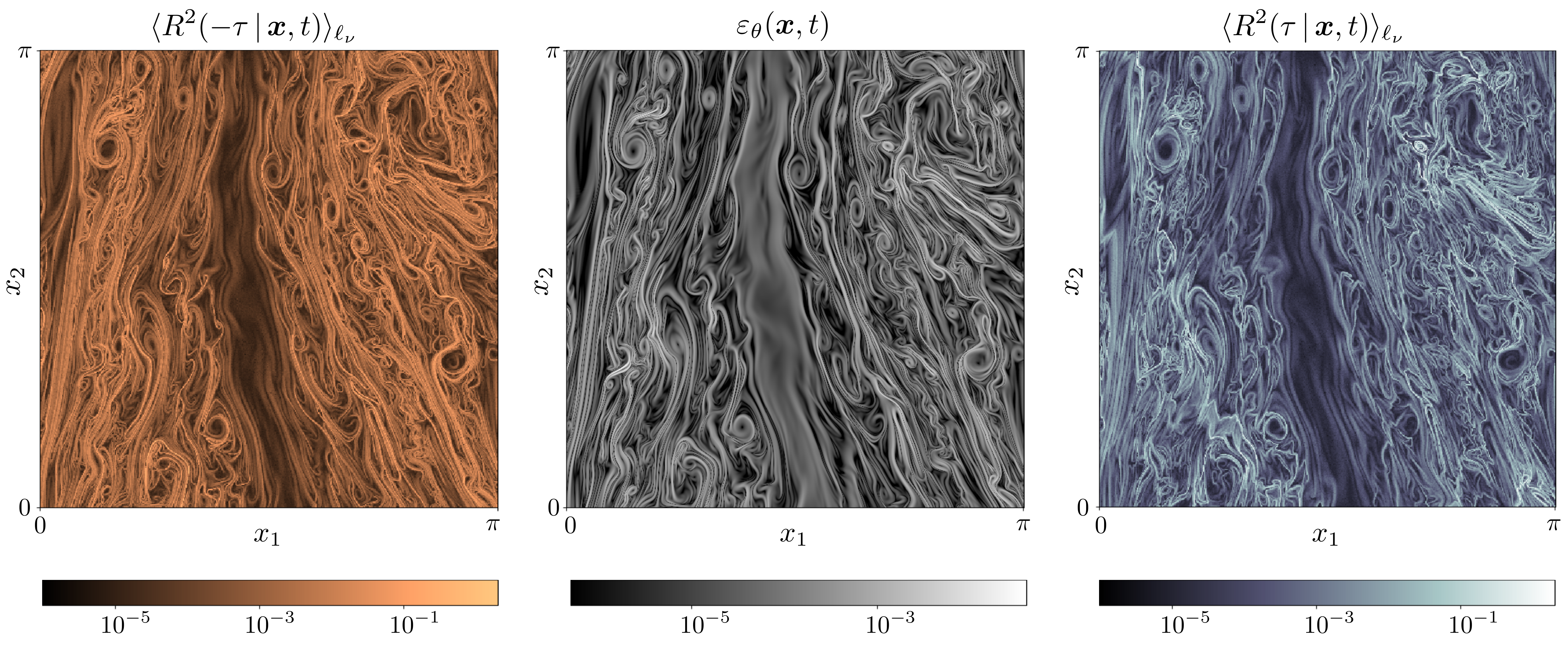}
\caption{\label{fig:triptych} 
{\it Left:} Backward mean-squared separation shown as a function of $\bx$ for $t=70$ and $\tau=5$ for Run IV.
{\it Center:} Dissipation field $\varepsilon_\theta(\bx,t) = \nu |\nabla\theta(\bx,t)|^2$.
{\it Right:} Forward mean-squared separation. See text for definitions.}
\end{figure}
The left and right panels of Fig.~\ref{fig:triptych} show the typical maps observed for the backward and forward relative separations computed from $t=70$ in the turbulent regime, and over a time lag $\tau = 5$. The central panel shows the dissipation field at the same instant of time. The correlation between these three fields is spectacular, with no apparent distinct features between forward and backward statistics. At a qualitative level, this is consistent with both the SQG fluctuation-dissipation relation (\ref{eq:fluct}) and the signature of incompressibility (\ref{eq:BF}).  Such a triptych representation substantiates the picture suggested by Fig.~(\ref{fig:dissip_anomal}) and the analysis of \S\ref{ssec:fieldstotraj}: SQG irreversibility implies Lagrangian spontaneous stochasticity.

\begin{figure}[h]
\centering
\includegraphics[width=\linewidth]{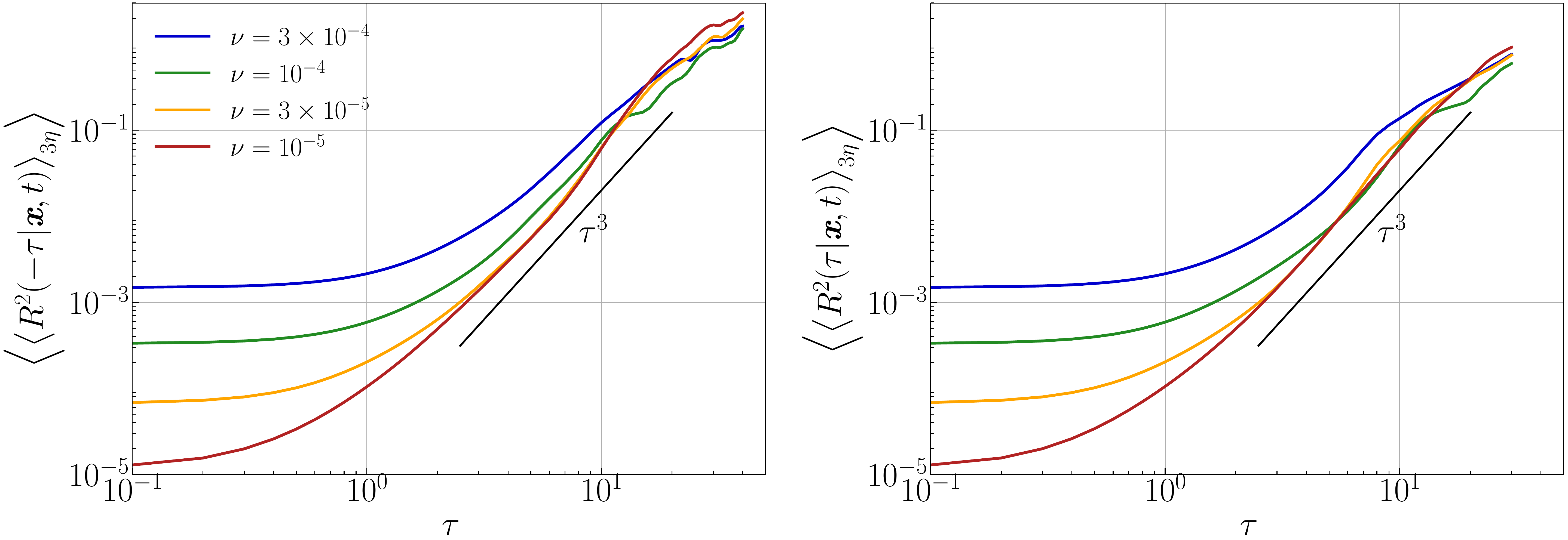}
\caption{\label{fig:Reynolds_FB_ts70}Mean-squared separations of pairs of tracers within puffs of sizes $\ell_\nu =3\eta$ at $t=70$, backward ({\it left}) and forward ({\it right}) in time, for various values of the viscosity, as labelled. The black lines show  behaviours $\propto \tau^3$ as expected from Richardson law.}
\end{figure}
For more quantitative assessments, we have plotted in Fig.~\ref{fig:Reynolds_FB_ts70} the backward and forward average relative separations for puffs of size $\ell_\nu= 3\eta$, at fixed $t =70$ and decreasing values of the viscosity.  In both cases, one observes a collapse of the statistics along a super-diffusive regime, compatible with Richardson's scaling  $\propto \tau^3$, although with possible deviations. As the viscosity decreases, the convergence towards Richardson's regime gets faster: This is a direct suggestive numerical evidence for the presence of Lagrangian spontaneous stochasticity,  characterised by the scaling regime extending down to $\tau=0^+$ in the joint limit  $\nu \to 0$ and $\ell_\nu\to 0$. In that limit, this signals finite-time separation of Lagrangian trajectories and non-deterministic nature for the limiting transition probabilities $\lim_{\nu\to 0}p_\nu$ ---\,provided that the identification (\ref{eq:mc}) indeed holds.

\begin{figure}[h]
\includegraphics[width=\linewidth]{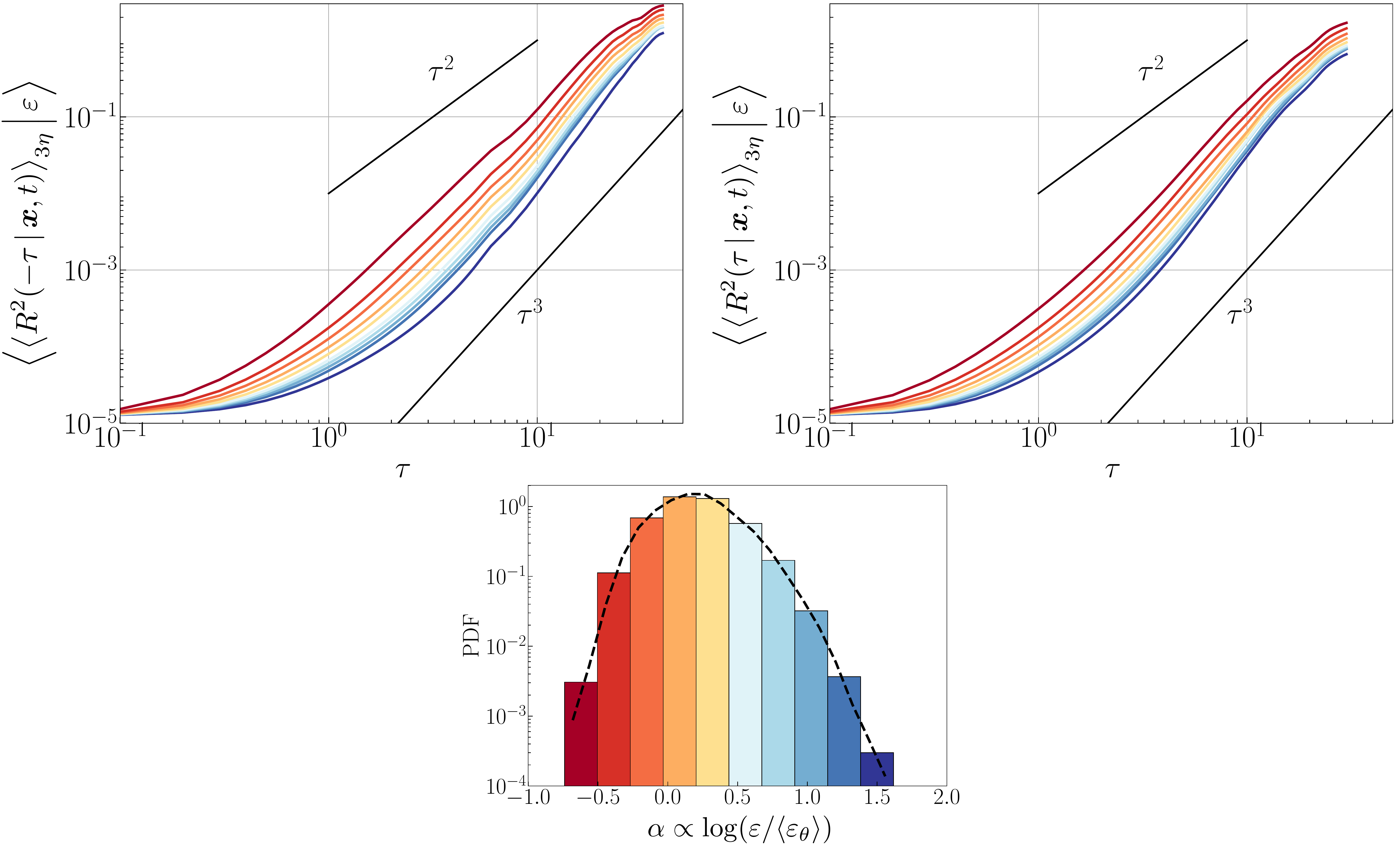}
\caption{\label{fig:Triptique_detail} Mean-squared separation conditioned on the local dissipation at $t= 70$, backward ({\it top left}) and forward ({\it top right}) in time. Different colours correspond to different values of $\varepsilon = \langle\varepsilon_\theta\rangle_{3\eta}$ as outlined on the histogram of $\alpha =  \log \left(\langle \varepsilon_\theta\rangle_\ell/\langle \varepsilon_\theta\rangle\right) / \log(\ell/L)$ in the bottom panel.}
\end{figure}

Besides spontaneous stochasticity, Figure~\ref{fig:Reynolds_FB_ts70} hardly reveals any additional sign of time irreversibility. In particular, there is no clear evidence that trajectories separate faster backward than forward in time. This is at variance with 3D Navier--Stokes turbulence, in which the backward mean-squared separation is observed to grow almost twice faster than the forward one~\cite{berg2006backwards,benveniste2014asymptotic,bourgoin2015turbulent,buaria2015characteristics,bragg2016forward}.  In our case, deterministic contributions, large-scale fluctuations, and possible intermittent corrections to Richardson's scaling prevent us from unambiguously estimating constants in front of the apparent $t^3$ scaling law. Nevertheless, the backward-forward asymmetry becomes more visible when further conditioning on the local dissipation rate $\av{\varepsilon_\theta (\bx,t)}_{\ell_\nu} = \varepsilon$.   We then write  $\left\langle\av{R^2(\pm \tau|\,\bx,t,\mu)}_{\ell_\nu}\,\middle|\,\varepsilon\right\rangle$ for the mean-squared separation from puffs with dissipation level $\varepsilon$. As seen in Fig.~\ref{fig:Triptique_detail}, pairs located in highly dissipative puffs (shown in red) have approached significantly faster than they subsequently separate. However, events that dissipate less than the average (in blue) show a slower dispersion backward than forward. This inverse tendency might compensate the bias due to dissipative events, possibly explaining the feeble manifestation of irreversibility in the total average of Fig.~\ref{fig:Reynolds_FB_ts70}.

\subsection{A tale of tempered stochasticity?}
\label{ssec:tempered}
\noindent Beyond the super-instability and breakdown of the Lagrangian flow evidenced in \S\ref{ssec:spontaneous}, another facet of spontaneous stochasticity relates to its  universality, and explicitly, to whether the limiting transition probabilities depend or not on the way the limit is taken. In the Kraichnan model, such a universality essentially holds when the advecting flow is incompressible: In that case, for example, the relative separation of tracers becomes independent of any small-scale regularisation of the underlying flow, provided separations are taken in the inertial range of scales.  For 3D Navier--Stokes turbulence, there are strong indications that such a scenario applies:  When initial separations are  larger than a few $\eta$,  backward and forward Lagrangian trajectories converge towards a universal scaling regime,
\begin{figure}[h]
\centering
\includegraphics[width=\linewidth]{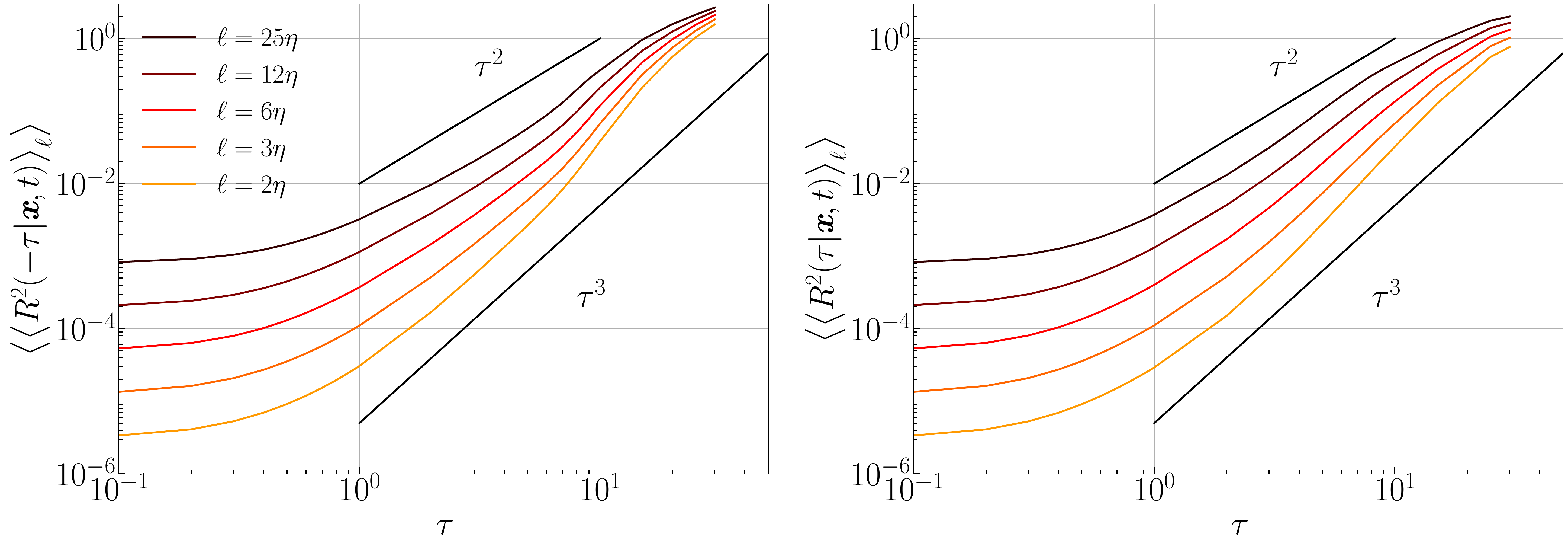}
\caption{\label{fig:dispersion_R0} 
Mean-squared separation of pairs of tracers within puffs of various sizes $\ell$ at $t=70$,  from  Run IV, backward ({\it left}) and forward ({\it right}) in time, with $\ell$ spanning the inertial range, as labelled.}
\end{figure}
$R^2(\pm\tau) \simeq g_\pm\varepsilon t^3$ with $g_+\approx g_-/2 \approx 0.55$ independent of $\eta$~\cite{bitane2012time, bourgoin2015turbulent,buaria2015characteristics,bragg2016forward}.
Figure~\ref{fig:dispersion_R0} indicates that, quite surprisingly, the stochasticity observed in our SQG deterministic setting is more \emph{tempered}:  The statistics of the relative separation $\langle R^2(\pm \tau\,|\,\bx,t) \rangle_\ell$ measured for pairs of travers within puffs of various inertial extensions $\ell$'s do not collapse towards a universal regime. Instead, the  anomalous diffusions have a non-universal prefactor that decreases with $\ell$. Such a non-universality is reminiscent of the weakly compressible phase observed in Kraichnan flows~\cite{falkovich2001particles, gawedzki2008soluble}, for which compressibility and roughness compensate each other, resulting in the small-scale behaviours of tracers trajectories fully determining their large-scale separations. As SQG flows are incompressible, the origin of such a behaviour is however not settled but could likely be due to non-trivial Lagrangian time correlations~\cite{chaves2003lagrangian}, and the deterministic nature of the large-scale flow.

\begin{figure}[h]
\centering
\includegraphics[width=\linewidth]{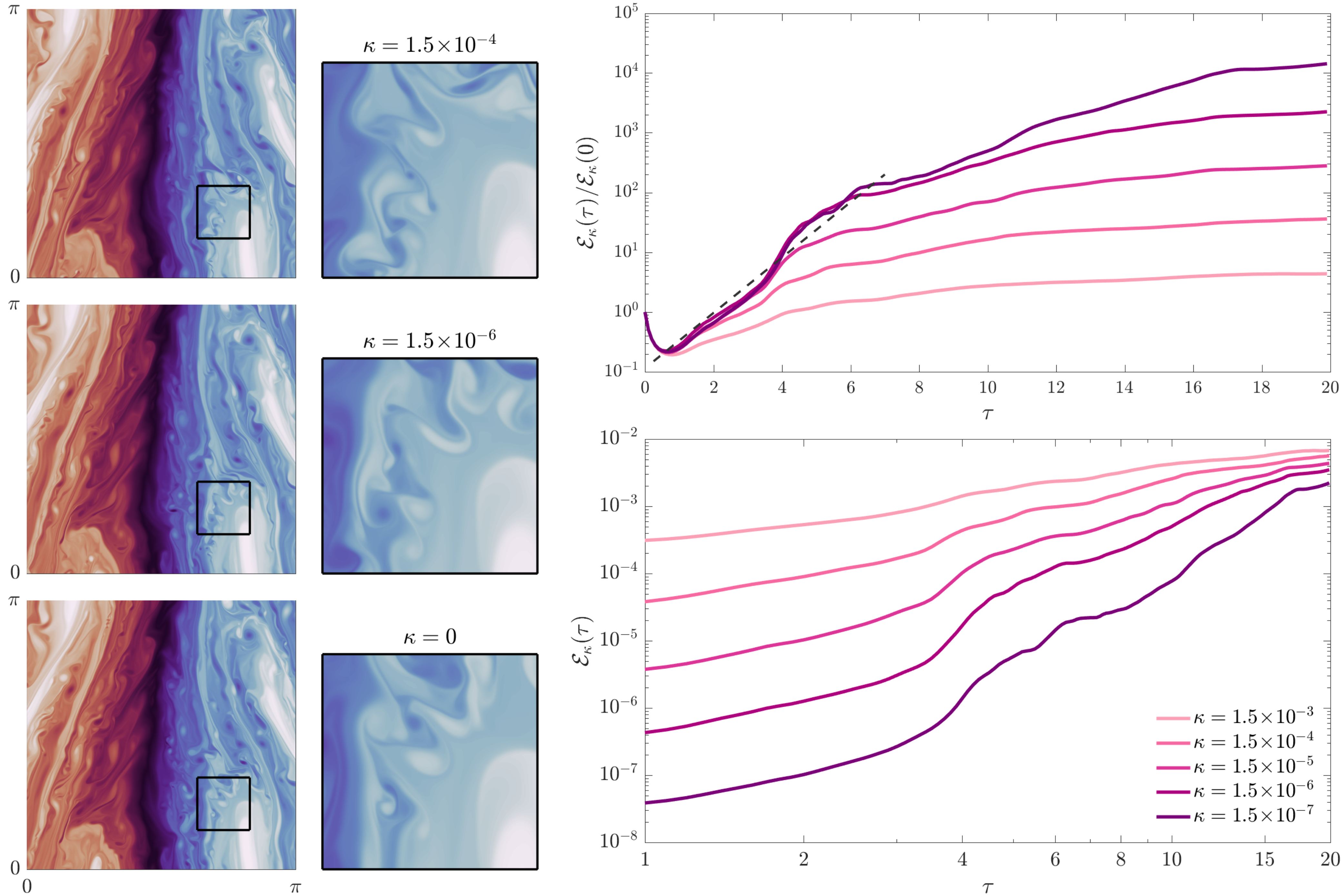}
\caption{\label{fig:crs_ss} Left: Snapshots at $t+\tau = 50$ of the solution $\theta^{(\kappa)}$ perturbed at $t=30$ with white noise of energy $\kappa \!= \!1.5\!\times\!10^{-4}$ ({\it top}) and $1.5\!\times\!10^{-6}$ ({\it middle}), together with the unperturbed case ({\it bottom}). The small panels show corresponding zooms in the region delimited by the black square. Right: Time evolution of the separation energy $\mathcal{E}_\kappa$ defined in (\ref{eq:sep_energ}) for various $\kappa$, with lin-log ({\it top}) and log-log ({\it bottom}) coordinates.}
\end{figure}

The same type of anomalous but non-universal behaviour is seen at the level of Eulerian predictability. Figure~\ref{fig:crs_ss} shows the time evolution of the separation energy 
\be
	\mathcal E_{\kappa} (\tau)= \frac{1}{2} \av{\left(\theta^{(\kappa)}(\cdot,t+\tau)- \theta(\cdot,t+\tau)\right)^2},
	\label{eq:sep_energ}
\ee
where the temperature field $\theta^{(\kappa)}$ is obtained from the reference field $\kappa$, by perturbing it at time $t$ with a spatial white-noise of amplitude $\sqrt{2\kappa}$, so that $\mathcal E_{\kappa} (0)= \kappa$. In the case of intrinsic stochasticity \emph{à la} Lorenz, one would expect the separation energy~(\ref{eq:sep_energ}) to display a universal scaling regime that becomes independent of $\kappa$ and extends towards $\tau = 0^+$ in the limit $\nu\to0$~\cite{thalabard2020butterfly}.  While for a fixed $\nu$, there is evidence for an initial chaotic, exponential stage $\mathcal E_{\kappa} (\tau) \simeq \kappa\, \mathrm{e}^{\lambda_\nu \tau}$, as well as for a later anomalous, algebraic separation regime, the latter is not universal, as both scaling and pre-factor depend upon the size $\kappa$ of the initial disturbance. Of course the lack of clear scaling law might be due to finite-size or viscosity effects. Still the lack of convergence towards a universal regime indicates that the intrinsic scenario \emph{\`a la} Lorenz might also be tempered, or even simply not realised. Indeed, our numerics cannot rule out that $\mathcal E_\kappa  \to 0$ as $\kappa$ and $\nu \to 0$. 

At a qualitative level, the snapshots in Fig.~\ref{fig:crs_ss} show that, while mesoscale features at $\ell \sim 2\pi/10$ and large-enough times differ for the various realisations, the large scales remain essentially undistinguishable. Our measurements contradict the scenario of a finite-time inverse cascade of errors that one would expect from simple models reproducing the $k^{-5/3}$ energy spectrum~\cite{lesieur1987turbulence,rotunno2008generalization}. As for Lagrangian statistics, the mechanism opposing the cascade of errors is unclear in the present setting. Possibilities include, but are not limited to, the deterministic nature of our setting, the specific choice of the initial condition $\theta_0$,
or the possible depletion of nonlinearity by a large-scale condensation.


\section{Concluding remarks}
\label{sec:conclusions}
\noindent Our highly-resolved numerics suggest that deterministic viscous SQG flows become turbulent upon decreasing the viscosity, with three hallmark signatures:  irreversibility, intermittency and spontaneous stochasticity of the Lagrangian flow.  Specific to the active-scalar nature of SQG is the fact that irreversibility of the advecting flow and Lagrangian stochasticity are two facets of the same phenomenon.  This is unlike the passive scalar case, where what matters is the irreversibility of the advected scalar, and also unlike Burgers, where this connection only holds at the level of the backward Lagrangian flow.  The present work also suggests that SQG spontaneous stochasticity is  \emph{tempered}, in the sense that the limiting stochastic flows may be highly sensitive to the details of the joint limit viscosity $\to$ 0, noise $\to$ 0, in both the Eulerian and the Lagrangian case.  Most limits could be deterministic, but spontaneous stochasticity may appear with ad-hoc choices. In a sense, this feature is reminiscent of the highly non-universal behaviour of tracers advected by weakly compressible random flows. It may signal a case where dependence upon initial condition could be neither differentiable (chaos), nor discontinuous (spontaneous stochasticity), but singular in between. However, it is unclear what physical mechanisms, which in SQG compete with roughness, are at play to temper stochasticity. Possibilities include the deterministic setup, or the presence of the Hamiltonian invariant, which favours large-scale deterministic self-organisation. For example, the predictability that is lost at mesoscales could be recovered from statistical averaging over turbulent scales. This would allow for randomness up to some intermediate mesoscale only, and not for the full dynamics. Investigating these possibilities are matters for future work.

\section*{Acknowledgements}
\noindent The authors acknowledge stimulating discussions with Mireille Bossy and Alexei Mailybaev. This work was supported by the French National Research Agency (ANR project TILT; ANR-20-CE30-0035). NV and JB  would like to thank the Isaac Newton Institute for Mathematical Sciences, Cambridge, for support and hospitality (through EPSRC grant no EP/R014604/1) during the program ``Mathematical aspects of turbulence: Where do we stand?'', where a part of this work was undertaken.

\bigskip
\bigskip
\noindent
Nicolas Valade,$^1$ Simon Thalabard,$^{1,2}$ and J\'er\'emie Bec$^{1,3}$\\[10pt]
e-mail:~\href{mailto:nicolas.valade@inria.fr}{nicolas.valade@inria.fr}\\
\phantom{email:~}\href{mailto:simon.thalabard@univ-cotedazur.fr}{simon.thalabard@univ-cotedazur.fr}\\
\phantom{email:~}\href{mailto:jeremie.bec@inria.fr}{jeremie.bec@inria.fr}\\[10pt]
$^1$~\inria\\[10pt]
$^2$~\inphyni\\[10pt]
$^3$~\cemef

\begin{thebibliography}{59}
\providecommand{\natexlab}[1]{#1}
\providecommand{\url}[1]{\texttt{#1}}
\expandafter\ifx\csname urlstyle\endcsname\relax
  \providecommand{\doi}[1]{doi: #1}\else
  \providecommand{\doi}{doi: \begingroup \urlstyle{rm}\Url}\fi

\bibitem[Kolmogorov(1941)]{kolmogorov1941local}
A.N. Kolmogorov.
\newblock The local structure of turbulence in incompressible viscous fluid for
  very large reynolds numbers.
\newblock \emph{Dokl. Akad. Nauk SSSR}, 30:\penalty0 301--305, 1941.
\newblock \doi{10.1098/rspa.1991.0075}.

\bibitem[Onsager(1949)]{onsager1949statistical}
L.~Onsager.
\newblock Statistical hydrodynamics.
\newblock \emph{Nuovo Cim.}, 6\penalty0 (2):\penalty0 279--287, 1949.
\newblock \doi{10.1007/BF02780991}.

\bibitem[De~Lellis(2017)]{de2017onsager}
C.~De~Lellis.
\newblock The {O}nsager theorem.
\newblock \emph{Surv. Diff. Geom.}, 22\penalty0 (1):\penalty0 71--101, 2017.

\bibitem[Isett(2018)]{isett2018proof}
P.~Isett.
\newblock A proof of {O}nsager's conjecture.
\newblock \emph{Ann. Math.}, 188\penalty0 (3):\penalty0 871--963, 2018.
\newblock \doi{10.4007/annals.2018.188.3.4}.

\bibitem[De~Lellis and Sz{\'e}kelyhidi~Jr(2012)]{delellis2012hprinciple}
C.~De~Lellis and L.~Sz{\'e}kelyhidi~Jr.
\newblock The $h$-principle and the equations of fluid dynamics.
\newblock \emph{Bull. Amer. Math. Soc.}, 49\penalty0 (3):\penalty0 347--375,
  2012.
\newblock \doi{10.1090/S0273-0979-2012-01376-9}.

\bibitem[Bardos et~al.(2014)Bardos, Sz{\'e}kelyhidi~Jr, and
  Wiedemann]{bardos2014nonuniqueness}
C.~Bardos, L.~Sz{\'e}kelyhidi~Jr, and E.~Wiedemann.
\newblock Non-uniqueness for the {E}uler equations: the effect of the boundary.
\newblock \emph{Russ. Math. Surv.}, 69\penalty0 (2):\penalty0 189, 2014.
\newblock \doi{10.1070/RM2014v069n02ABEH004886}.

\bibitem[Frisch(1995)]{frisch1995turbulence}
U.~Frisch.
\newblock \emph{Turbulence: the legacy of {A}.{N}. {K}olmogorov}.
\newblock Cambridge University Press, Cambridge, UK, 1995.

\bibitem[Kolmogorov(1962)]{kolmogorov1962refinement}
A.N. Kolmogorov.
\newblock A refinement of previous hypotheses concerning the local structure of
  turbulence in a viscous incompressible fluid at high {R}eynolds number.
\newblock \emph{J. Fluid Mech.}, 13\penalty0 (1):\penalty0 82--85, 1962.
\newblock \doi{10.1017/S0022112062000518}.

\bibitem[Chevillard et~al.(2019)Chevillard, Garban, Rhodes, and
  Vargas]{chevillard2019skewed}
L.~Chevillard, C.~Garban, R.~Rhodes, and V.~Vargas.
\newblock On a skewed and multifractal unidimensional random field, as a
  probabilistic representation of {K}olmogorov’s views on turbulence.
\newblock \emph{Ann. Henri Poincar\'{e}}, 20\penalty0 (11):\penalty0
  3693--3741, 2019.
\newblock \doi{10.1007/s00023-019-00842-y}.

\bibitem[Mailybaev and Thalabard(2022)]{mailybaev2022hidden}
A.A. Mailybaev and S.~Thalabard.
\newblock Hidden scale invariance in {N}avier--{S}tokes intermittency.
\newblock \emph{Phil. Trans. Roy. Soc.}, 380\penalty0 (2218):\penalty0
  20210098, 2022.
\newblock \doi{10.1098/rsta.2021.0098}.

\bibitem[Gaw\k{e}dzki(2001)]{gawedzki2001turbulent}
K.~Gaw\k{e}dzki.
\newblock Turbulent advection and breakdown of the {L}agrangian flow.
\newblock In \emph{Intermittency in Turbulent Flows}, Edited by Vassilicos,
  J.C., pages 86--104. Cambridge University Press, 2001.

\bibitem[Gaw\k{e}dzki(2006)]{gawedzki2006simple}
K.~Gaw\k{e}dzki.
\newblock Simple models of turbulent transport.
\newblock In \emph{XIVth International Congress On Mathematical Physics}, pages
  38--49. World Scientific, 2006.

\bibitem[Gaw\k{e}dzki(2008)]{gawedzki2008soluble}
K.~Gaw\k{e}dzki.
\newblock Soluble models of turbulent transport.
\newblock In \emph{Non-equilibrium statistical mechanics and turbulence},
  number 355 in Lond. Math. Soc. Lect. edited by Nazarenko, S. and Zaboronski,
  O., chapter~2. Cambridge University Press, 2008.

\bibitem[Bernard et~al.(1998)Bernard, Gaw\k{e}dzki, and
  Kupiainen]{bernard1998slow}
D.~Bernard, K.~Gaw\k{e}dzki, and A.~Kupiainen.
\newblock Slow modes in passive advection.
\newblock \emph{J. Stat. Phys.}, 90\penalty0 (3):\penalty0 519--569, 1998.
\newblock \doi{10.1023/A:1023212600779}.

\bibitem[Falkovich et~al.(2001)Falkovich, Gaw\k{e}dzki, and
  Vergassola]{falkovich2001particles}
G.~Falkovich, K.~Gaw\k{e}dzki, and M.~Vergassola.
\newblock Particles and fields in fluid turbulence.
\newblock \emph{Rev. Mod. Phys.}, 73:\penalty0 913--975, 2001.
\newblock \doi{10.1103/RevModPhys.73.913}.

\bibitem[Kupiainen(2003)]{kupiainen2003nondeterministic}
A~Kupiainen.
\newblock Nondeterministic dynamics and turbulent transport.
\newblock \emph{Ann. Henri Poincar\'{e}}, 4\penalty0 (2):\penalty0 713--726,
  2003.
\newblock \doi{10.1007/s00023-003-0957-3}.

\bibitem[E and Vanden~Eijnden(2000)]{e2000generalized}
W.~E and E.~Vanden~Eijnden.
\newblock Generalized flows, intrinsic stochasticity, and turbulent transport.
\newblock \emph{Proc. Nat. Acad. Sci. U.S.A.}, 97:\penalty0 8200--8205, 2000.
\newblock \doi{10.1073/pnas.97.15.8200}.

\bibitem[Chaves et~al.(2003)Chaves, Gaw\k{e}dzki, Horvai, Kupiainen, and
  Vergassola]{chaves2003lagrangian}
M.~Chaves, K.~Gaw\k{e}dzki, P.~Horvai, A.~Kupiainen, and M.~Vergassola.
\newblock Lagrangian dispersion in {G}aussian self-similar velocity ensembles.
\newblock \emph{J. Stat. Phys.}, 113\penalty0 (5):\penalty0 643--692, 2003.
\newblock \doi{10.1023/A:1027348316456}.

\bibitem[Le~Jan and Raimond(2002)]{le2002integration}
Y.~Le~Jan and O.~Raimond.
\newblock Integration of {B}rownian vector fields.
\newblock \emph{Ann. Probab.}, 30\penalty0 (2):\penalty0 826--873, 2002.
\newblock \doi{10.1214/aop/1023481009}.

\bibitem[Drivas and Mailybaev(2021)]{drivas2021life}
T.D. Drivas and A.A. Mailybaev.
\newblock ‘{L}ife after death’ in ordinary differential equations with a
  non-{L}ipschitz singularity.
\newblock \emph{Nonlinearity}, 34:\penalty0 2296--2326, 2021.
\newblock \doi{10.1088/1361-6544/abbe60}.

\bibitem[Salazar and Collins(2009)]{salazar2009two}
J.P.L.C. Salazar and L.R. Collins.
\newblock Two-particle dispersion in isotropic turbulent flows.
\newblock \emph{Annu. Rev. Fluid Mech.}, 41\penalty0 (1):\penalty0 405--432,
  2009.
\newblock \doi{10.1146/annurev.fluid.40.111406.102224}.

\bibitem[Lorenz(1969)]{lorenz69predictability}
E.~Lorenz.
\newblock The predictability of a flow which possesses many scales of motion.
\newblock \emph{Tellus}, 21\penalty0 (3):\penalty0 289--307, 1969.
\newblock \doi{10.3402/tellusa.v21i3.10086}.

\bibitem[Palmer et~al.(2014)Palmer, D{\"o}ring, and Seregin]{palmer2014real}
T.~Palmer, A.~D{\"o}ring, and G.~Seregin.
\newblock The real butterfly effect.
\newblock \emph{Nonlinearity}, 27\penalty0 (9):\penalty0 R123, 2014.
\newblock \doi{10.1088/0951-7715/27/9/R123}.

\bibitem[Mailybaev(2016)]{mailybaev2016spontaneous}
A.A. Mailybaev.
\newblock Spontaneous stochasticity of velocity in turbulence models.
\newblock \emph{Multiscale Model. Simul.}, 14\penalty0 (1):\penalty0 96--112,
  2016.
\newblock \doi{10.1137/15M1012451}.

\bibitem[Mailybaev and Raibekas(2022)]{mailybaev2022spontaneous}
A.A. Mailybaev and A.~Raibekas.
\newblock Spontaneous stochasticity and renormalization group in discrete
  multi-scale dynamics.
\newblock \emph{arXiv preprint arXiv:2207.06158}, 2022.

\bibitem[Biferale et~al.(2018)Biferale, Boffetta, Mailybaev, and
  Scagliarini]{biferale2018rayleigh-taylor}
L.~Biferale, G.~Boffetta, A.A. Mailybaev, and A.~Scagliarini.
\newblock Rayleigh-{T}aylor turbulence with singular nonuniform initial
  conditions.
\newblock \emph{Phys. Rev. Fluids}, 3\penalty0 (9):\penalty0 092601, 2018.
\newblock \doi{10.1103/PhysRevFluids.3.092601}.

\bibitem[Thalabard et~al.(2020)Thalabard, Bec, and
  Mailybaev]{thalabard2020butterfly}
S.~Thalabard, J.~Bec, and A.A. Mailybaev.
\newblock From the butterfly effect to spontaneous stochasticity in singular
  shear flows.
\newblock \emph{Commun. Phys.}, 3:\penalty0 122, 2020.
\newblock \doi{10.1038/s42005-020-0391-6}.

\bibitem[Frisch and Bec(2001)]{frisch2001burgulence}
U.~Frisch and J.~Bec.
\newblock Burgulence.
\newblock In \emph{New trends in turbulence}, pages 341--383. Springer, 2001.

\bibitem[Eyink and Drivas(2015)]{eyink2015spontaneous}
G.L. Eyink and T.D. Drivas.
\newblock Spontaneous stochasticity and anomalous dissipation for {B}urgers
  equation.
\newblock \emph{J. Stat. Phys.}, 158:\penalty0 386--432, 2015.
\newblock \doi{10.1007/s10955-014-1135-3}.

\bibitem[Drivas and Eyink(2017)]{drivas2017lagrangian}
T.D. Drivas and G.L. Eyink.
\newblock A {L}agrangian fluctuation--dissipation relation for scalar
  turbulence. {P}art {I}. {F}lows with no bounding walls.
\newblock \emph{J. Fluid Mech.}, 829:\penalty0 153--189, 2017.
\newblock \doi{10.1017/jfm.2017.567}.

\bibitem[Blumen(1978)]{blumen1978uniform}
W.~Blumen.
\newblock Uniform potential vorticity flow: {P}art {I}. {T}heory of wave
  interactions and two-dimensional turbulence.
\newblock \emph{J. Atmos. Sci.}, 35\penalty0 (5):\penalty0 774--783, 1978.
\newblock \doi{10.1175/1520-0469(1978)035<0774:UPVFPI>2.0.CO;2}.

\bibitem[Held et~al.(1995)Held, Pierrehumbert, Garner, and
  Swanson]{held1995surface}
I.~Held, R.~Pierrehumbert, S.~Garner, and K.~Swanson.
\newblock Surface quasi-geostrophic dynamics.
\newblock \emph{J. Fluid Mech.}, 282:\penalty0 1--20, 1995.
\newblock \doi{10.1017/S0022112095000012}.

\bibitem[Lapeyre(2017)]{lapeyre2017surface}
G.~Lapeyre.
\newblock Surface quasi-geostrophy.
\newblock \emph{Fluids}, 2:\penalty0 7, 2017.
\newblock \doi{10.3390/fluids2010007}.

\bibitem[Constantin et~al.(1994)Constantin, Majda, and
  Tabak]{constantin1994formation}
P.~Constantin, A.J. Majda, and E.~Tabak.
\newblock Formation of strong fronts in the {2D} quasigeostrophic thermal
  active scalar.
\newblock \emph{Nonlinearity}, 7:\penalty0 1495--1533, 1994.
\newblock \doi{10.1088/0951-7715/7/6/001}.

\bibitem[Constantin and Wu(1999)]{constantin1999behavior}
P.~Constantin and J.~Wu.
\newblock Behavior of solutions of {2D} quasi-geostrophic equations.
\newblock \emph{SIAM J. Math. Anal.}, 30\penalty0 (5):\penalty0 937--948, 1999.
\newblock \doi{10.1137/S0036141098337333}.

\bibitem[Buckmaster et~al.(2019)Buckmaster, Shkoller, and
  Vicol]{buckmaster2019nonuniqueness}
T.~Buckmaster, S.~Shkoller, and V.~Vicol.
\newblock Nonuniqueness of weak solutions to the {SQG} equation.
\newblock \emph{Commun. Pure Appl. Math.}, 72:\penalty0 1809--1874, 2019.
\newblock \doi{10.1002/cpa.21851}.

\bibitem[Smith et~al.(2002)Smith, Boccaletti, Henning, Marinov, Tam, Held, and
  Vallis]{smith2002turbulent}
K.~Smith, G.~Boccaletti, C.~Henning, I.~Marinov, C.~Tam, I.~Held, and
  G.~Vallis.
\newblock Turbulent diffusion in the geostrophic inverse cascade.
\newblock \emph{J. Fluid Mech.}, 469:\penalty0 13--48, 2002.
\newblock \doi{10.1017/S0022112002001763}.

\bibitem[Celani et~al.(2004)Celani, Cencini, Mazzino, and
  Vergassola]{celani2004active}
A.~Celani, M.~Cencini, A.~Mazzino, and M.~Vergassola.
\newblock Active and passive fields face to face.
\newblock \emph{New J. Phys.}, 6:\penalty0 72, 2004.
\newblock \doi{10.1088/1367-2630/6/1/072}.

\bibitem[Rotunno and Snyder(2008)]{rotunno2008generalization}
R.~Rotunno and C.~Snyder.
\newblock A generalization of {L}orenz’s model for the predictability of
  flows with many scales of motion.
\newblock \emph{J. Atmos. Sci.}, 65\penalty0 (3):\penalty0 1063--1076, 2008.
\newblock \doi{10.1175/2007JAS2449.1}.

\bibitem[Foussard et~al.(2017)Foussard, Berti, Perrot, and
  Lapeyre]{foussard2017relative}
A.~Foussard, S.~Berti, X.~Perrot, and G.~Lapeyre.
\newblock Relative dispersion in generalized two-dimensional turbulence.
\newblock \emph{J. Fluid Mech.}, 821:\penalty0 358--383, 2017.
\newblock \doi{10.1017/jfm.2017.253}.

\bibitem[Sukhatme and Pierrehumbert(2002)]{sukhatme2002surface}
J.~Sukhatme and R.T. Pierrehumbert.
\newblock Surface quasigeostrophic turbulence: {T}he study of an active scalar.
\newblock \emph{Chaos}, 12\penalty0 (2):\penalty0 439--450, 2002.
\newblock \doi{10.1063/1.1480442}.

\bibitem[Kiselev(2020)]{kiselev2020small}
A.~A. Kiselev.
\newblock Small scale creation in active scalars.
\newblock In L.C. Berselli and M.~R{\r{u}}{\v{z}}i{\v{c}}ka, editors,
  \emph{Progress in Mathematical Fluid Dynamics}, pages 125--161, Cetraro,
  Italy, 2020. Springer.
\newblock \doi{10.1007/978-3-030-54899-5}.

\bibitem[Ohkitani and Yamada(1997)]{ohkitani1997inviscid}
K.~Ohkitani and M.~Yamada.
\newblock Inviscid and inviscid-limit behavior of a surface quasigeostrophic
  flow.
\newblock \emph{Phys. Fluids}, 9\penalty0 (4):\penalty0 876--882, 1997.
\newblock \doi{10.1063/1.869184}.

\bibitem[Constantin et~al.(1998)Constantin, Nie, and
  Sch{\"o}rghofer]{constantin1998nonsingular}
P.~Constantin, Q.~Nie, and N.~Sch{\"o}rghofer.
\newblock Nonsingular surface quasi-geostrophic flow.
\newblock \emph{Phys. Lett. A}, 241\penalty0 (3):\penalty0 168--172, 1998.
\newblock \doi{10.1016/S0375-9601(98)00108-X}.

\bibitem[Cordoba(1998)]{cordoba1998nonexistence}
D.~Cordoba.
\newblock Nonexistence of simple hyperbolic blow-up for the quasi-geostrophic
  equation.
\newblock \emph{Ann. Math.}, 148\penalty0 (3):\penalty0 1135--1152, 1998.
\newblock \doi{10.2307/121037}.

\bibitem[Scott and Dritschel(2019)]{scott2019scale}
R.~K. Scott and D.~G. Dritschel.
\newblock Scale-invariant singularity of the surface quasigeostrophic patch.
\newblock \emph{J. Fluid Mech.}, 863:\penalty0 R2, 2019.
\newblock \doi{10.1017/jfm.2019.7}.

\bibitem[Resnick(1995)]{resnick1995dynamical}
S.G. Resnick.
\newblock \emph{Dynamical problems in non-linear advective partial differential
  equations}.
\newblock PhD thesis, PhD Thesis, The University of Chicago, 1995.

\bibitem[Isett and Vicol(2015)]{isett2015holder}
P.~Isett and V.~Vicol.
\newblock H{\"o}lder continuous solutions of active scalar equations.
\newblock \emph{Ann. PDE}, 1\penalty0 (1):\penalty0 1--77, 2015.
\newblock \doi{10.1007/s40818-015-0002-0}.

\bibitem[Akramov and Wiedemann(2019)]{akramov2019renormalization}
I.~Akramov and E.~Wiedemann.
\newblock Renormalization of active scalar equations.
\newblock \emph{Nonlinear Anal.}, 179:\penalty0 254--269, 2019.
\newblock \doi{10.1016/j.na.2018.08.018}.

\bibitem[Nelkin(1990)]{nelkin1990multifractal}
M.~Nelkin.
\newblock Multifractal scaling of velocity derivatives in turbulence.
\newblock \emph{Phys. Rev. A}, 42\penalty0 (12):\penalty0 7226--7229, 1990.
\newblock \doi{10.1103/PhysRevA.42.7226}.

\bibitem[Constantin and Iyer(2008)]{constantin2008stochastic}
P.~Constantin and G.~Iyer.
\newblock A stochastic {L}agrangian representation of the three-dimensional
  incompressible {N}avier-{S}tokes equations.
\newblock \emph{Commun. Pure Appl. Math.}, 61:\penalty0 330--345, 2008.
\newblock \doi{10.1002/cpa.20192}.

\bibitem[Evans(2012)]{evans2012introduction}
L.~Evans.
\newblock \emph{An introduction to stochastic differential equations},
  volume~82.
\newblock Amer. Math. Soc., 2012.

\bibitem[Berg et~al.(2006)Berg, L{\"u}thi, Mann, and Ott]{berg2006backwards}
J.~Berg, B.~L{\"u}thi, J.~Mann, and S.~Ott.
\newblock Backwards and forwards relative dispersion in turbulent flow: an
  experimental investigation.
\newblock \emph{Phys. Rev. E}, 74\penalty0 (1):\penalty0 016304, 2006.
\newblock \doi{10.1103/PhysRevE.74.016304}.

\bibitem[Benveniste and Drivas(2014)]{benveniste2014asymptotic}
D.~Benveniste and T.~Drivas.
\newblock Asymptotic results for backwards two-particle dispersion in a
  turbulent flow.
\newblock \emph{Phys. Rev. E}, 89\penalty0 (4):\penalty0 041003, 2014.
\newblock \doi{10.1103/PhysRevE.89.041003}.

\bibitem[Bourgoin(2015)]{bourgoin2015turbulent}
M.~Bourgoin.
\newblock Turbulent pair dispersion as a ballistic cascade phenomenology.
\newblock \emph{J. Fluid Mech.}, 772:\penalty0 678--704, 2015.
\newblock \doi{10.1017/jfm.2015.206}.

\bibitem[Buaria et~al.(2015)Buaria, Sawford, and
  Yeung]{buaria2015characteristics}
D.~Buaria, B.~Sawford, and P-K. Yeung.
\newblock Characteristics of backward and forward two-particle relative
  dispersion in turbulence at different reynolds numbers.
\newblock \emph{Phys. Fluids}, 27\penalty0 (10):\penalty0 105101, 2015.
\newblock \doi{10.1063/1.4931602}.

\bibitem[Bragg et~al.(2016)Bragg, Ireland, and Collins]{bragg2016forward}
A.~Bragg, P.~Ireland, and L.~Collins.
\newblock Forward and backward in time dispersion of fluid and inertial
  particles in isotropic turbulence.
\newblock \emph{Phys. Fluids}, 28\penalty0 (1):\penalty0 013305, 2016.
\newblock \doi{10.1063/1.4939694}.

\bibitem[Bitane et~al.(2012)Bitane, Homann, and Bec]{bitane2012time}
R.~Bitane, H.~Homann, and J.~Bec.
\newblock Time scales of turbulent relative dispersion.
\newblock \emph{Phys. Rev. E}, 86\penalty0 (4):\penalty0 045302, 2012.
\newblock \doi{10.1103/PhysRevE.86.045302}.

\bibitem[Lesieur(1987)]{lesieur1987turbulence}
M.~Lesieur.
\newblock \emph{Turbulence in fluids: stochastic and numerical modelling},
  volume 488.
\newblock Nijhoff Boston, MA, 1987.

\end{thebibliography}
\end{document}